\documentclass[graybox]{svmult}

\usepackage{mathptmx}       
\usepackage{helvet}         
\usepackage{courier}        
\usepackage{type1cm}        
\usepackage{makeidx}         
\usepackage{graphicx}        
\usepackage{multicol}        
\usepackage[bottom]{footmisc}

\makeindex             

\begin{document}

\title*{Evolutionary changes in the periods \\of galactic RRab stars}
\titlerunning{Evolutionary changes in RRab stars } 
\author{Vandenbroere, J., Poretti, E., and Le Borgne, J.F.}
\authorrunning{Vandenbroere et al.} 
\institute{Vandenbroere, Jacqueline \at GEOS,  23 Parc de Levesville, 28300 Bailleau l'Eveque, France.
 \email{j.vandenbroere@skynet.be}
\and Poretti, Ennio \at INAF-OA Brera, Via E. Bianchi 46, 23879 Merate, Italy. \email{ennio.poretti@brera.inaf.it}
\and Le Borgne, Jean-Francois \at Institut de Recherche en Astrophysique et Plan\'etologie, 14 Avenue E. Belin,
31400 Toulouse, France. \email{jean-francois.leborgne@irap.omp.eu}  }
%
%
\maketitle

\abstract*{Le Borgne et al. (2007) report on the determination of evolutionary changes in
the periods of field RR Lyr stars. Thanks to the extension of the GEOS database, we could 
analyze a sample twice larger than the previous one. We obtained a different picture of the
period changes, with a number of stars showing an increasing period greater than
that of stars showing a decreasing period.}
\abstract{Le Borgne et al. (2007) report on the determination of evolutionary changes in
the periods of field RR Lyr stars. Thanks to the extension of the GEOS database, we could 
analyze a sample twice larger than the previous one. We obtained a different picture of the
period changes, with a number of stars showing an increasing period twice greater than
that of stars showing a decreasing period.}

\section*{}
The amateur/professional association GEOS (Groupe Europ\'een d'Observations
Stellaires) built a database aimed to put together the times  of maximum light
of RR Lyr stars  published in the literature, coming back to end of XIX$^{\rm th}$ century.
The analysis of the differences between the {\it observed} and {\it calculated} times
of maximum brightness (O-C values)  over a timescale
of more than 100 years is one of the few tool able to provide quantitative
tests of the stellar evolution theory.
The GEOS database is freely accessible on the internet at the address
{\tt http://rr-lyr.ast.obs-mip.fr/}. 

Le Borgne at al. (2007) analyzed 123 galactic RRab stars showing a clear O--C pattern
(constant, parabolic or erratic) and  
found clear evidence of period increases or decreases at constant rates,
suggesting evolutionary effects.
One of the most interesting results  was that RRab stars showing blueward evolution 
(i.e., period decreases) are quite common, slightly less than RRab stars 
showing redward evolution (i.e., period increases).
The number of maxima of RRab stars in the GEOS database increased rapidly in number and
precision thanks to the survey performed with the TAROT telescopes and by amateur astronomers
equipped with CCDs.
Therefore, a few years later we could double the number of with obvious period variations 
\cite{geos}.

To date, we have studied the evolution of the period of 217 RRab stars for which at 
least 20 times of maximum spanning 50 years are available. Table~\ref{change} 
summarizes the results. 
The incidence of stars with decreasing periods is still relevant and we could
confirm that the
blueward  path has a non-negligible part in the evolution of horizontal branch
stars. The new fact is that  
the number of RRab stars with increasing periods 
is now clearly larger than that of stars with decreasing periods. This partially
solve one of the problems raised in the previous analysis, i.e., the necessity
to invoke a fast redward evolution to explain the   small ratio between the number
of stars with increasing period and that of stars with decreasing period.

We note that this statistics
is based on stars showing a constant rate, which could be 
determined in  a very reliable way and then straightly referred to evolutionary effects. 
Erratic or irregular changes are more
difficult to be put in a precise evolutionary scenario.

\begin{table}
\begin{center}
\caption{Statistics of RRab stars showing period changes due to evolutionary effects}
\label{change}       
%
%
\begin{tabular}{ll clclc}
\hline\noalign{\smallskip}
\multicolumn{1}{c}{Type of } && \multicolumn{1}{c}{Number of} && 
\multicolumn{1}{c}{Percentage} && \multicolumn{1}{c}{Mean dP/dt} \\ 
\multicolumn{1}{c}{Period} && \multicolumn{1}{c}{stars} && 
\multicolumn{1}{c}{[\%]} && \multicolumn{1}{c}{[10$^{-10}$\,d/d]} \\ 
\noalign{\smallskip}\svhline\noalign{\smallskip}
Constant && 104  && 48 &&  \\
Increasing && 56  && 26 && +6.83 \\
Decreasing& & 28  && 13 && -10.53 \\
Irregular && 29   && 13 &&  \\
\noalign{\smallskip}\hline\noalign{\smallskip}
\end{tabular}
\end{center}
\end{table}

\begin{acknowledgement}
EP acknowledges financial support from the {\it Institut de Recherche en Astrophysique 
et Plan\'etologie} during his stay in Toulouse, where part of this work has
been prepared.
\end{acknowledgement}

\end{document}